\documentclass[%
 amsmath,amssymb,
 aps, b
 prl,
 twocolumn,
 superscriptaddress,
 floatfix
]{revtex4-2}

\usepackage{graphicx}
\usepackage{dcolumn}
\usepackage{bm}
\usepackage{color}
\usepackage{soul}
\usepackage{ulem}
\usepackage{hyperref}

\begin{document}

\preprint{APS/123-QED}

\title{Stochastic jetting and dripping in confined soft granular flows}

\author{Micha{\l} Bogdan}
\email[]{mbogdan@ichf.edu.pl}
\affiliation{Institute of Physical Chemistry, Polish Academy of Sciences, Kasprzaka 44/52,
01-224 Warsaw, Poland}

\author{Andrea Montessori}
\affiliation{Dipartimento di Ingegneria, Università degli Studi Roma tre, via Vito Volterra 62, Rome, 00146, Italy}

\author{Adriano Tiribocchi}
\affiliation{Istituto per le Applicazioni del Calcolo del Consiglio Nazionale delle Ricerche, via dei Taurini 19, 00185, Rome, Italy}

\author{Fabio Bonaccorso}
\affiliation{Istituto per le Applicazioni del Calcolo del Consiglio Nazionale delle Ricerche, via dei Taurini 19, 00185, Rome, Italy}
\affiliation{Department of Physics and National Institute for Nuclear Physics, University of Rome "Tor Vergata'', Via Cracovia, 50, 00133 Rome, Italy}

\author{Marco Lauricella}
\affiliation{Istituto per le Applicazioni del Calcolo del Consiglio Nazionale delle Ricerche, via dei Taurini 19, 00185, Rome, Italy}

\author{Leon Jurkiewicz}
\affiliation{Institute of Physical Chemistry, Polish Academy of Sciences, Kasprzaka 44/52,
01-224 Warsaw, Poland}

\author{Sauro Succi}
\affiliation{Istituto per le Applicazioni del Calcolo del Consiglio Nazionale delle Ricerche, via dei Taurini 19, 00185, Rome, Italy}
\affiliation{Center for Life Nanoscience at la Sapienza, Istituto Italiano di Tecnologia, viale Regina Elena 295, 00161, Rome, Italy}
\affiliation{Department of Physics, Harvard University, 17 Oxford St, Cambridge, MA 02138, United States}

\author{Jan Guzowski}
\email[]{jguzowski@ichf.edu.pl}
\affiliation{Institute of Physical Chemistry, Polish Academy of Sciences, Kasprzaka 44/52,
01-224 Warsaw, Poland}

\date{\today}

\begin{abstract}

We report new dynamical modes in confined soft granular flows, such as 
stochastic jetting and dripping, with no counterpart in continuum viscous fluids.
The new modes emerge as a result of the propagation of the chaotic behaviour of individual grains- here, monodisperse emulsion droplets- to the level of the entire system as the emulsion is focused into a narrow orifice by an external viscous flow. 
We observe avalanching dynamics and the formation of remarkably stable jets- single-file granular chains- which occasionally break, resulting in a non-Gaussian distribution of cluster sizes. We find that the sequences of droplet rearrangements that lead to 
the formation of such chains resemble unfolding
of cancer cell clusters in narrow capillaries, overall demonstrating that microfluidic emulsion systems could serve to model various aspects of soft granular flows, including also tissue dynamics at the meso-scale.

\end{abstract}

\maketitle



Soft granular materials consist of close-packed deformable grains separated by thin fluid films. 
They are ubiquitous in industries, forming food and cosmetic products and in nature, examples 
including dense emulsions, foams, as well as certain types of biological tissues, among others \cite{Guevorkian2010, Douezan2011, Manning2010, Cohen-Addad2013, Nezamabadi2017, Kabla2012, Pawlizak2015}. 
The presence of the internal lengthscale in such materials, associated with the grain size, leads to a 
complex many-body dynamics governed by the sequences of grain deformations 
and rearrangements \cite{Gai2016, Gai2016a}, which in turn result in complex flows and rheological behavior, including 
plasticity and viscoelasticity, memory effects and avalanches \cite{Jiang1999, Marmottant2013, Kumar2020, Goyon2008, Goyon2010, Lulli2018}.

The flow of such types of materials confined to narrow geometries is of primary interest to 
the physics of amorphous solids and glasses \cite{Uchic2004, Gai2016, Gai2016a},  as well 
as of technological relevance for the generation of compartmentalized capsules \cite{Constantini2018} 
and porous materials \cite{Constantini2019} or in bioprinting \cite{Highley2019}. 
The dynamics of soft granular media in constrictions under external flow is also of significant 
interest in tissue mechanics \cite{Au2016}, as it could shed light on the behavior of cell clusters 
passing through physiological constrictions, a process that remains one of the critical stages 
of tumor metastasis.

Previous microfludic approaches to soft granular materials addressed the behavior of foams or dense emulsions inside channels, however without considering the interaction with an external flow \cite{Raven2009, Garstecki2006, Gai2016, Gai2016a, Goyon2008}. Here, we systematically study the behavior of a model 'wet' soft-granular medium (a tightly packed monodisperse emulsion) under external viscous forces. We use a flow-focusing geometry in which the emulsion is fed through the middle channel and the external immiscible phase through the side channels. Accordingly, the emulsion is focused by the external flow and narrows until passing through an orifice, a situation which closely resembles the flow of cell clusters in capillaries.

Typically, in microfluidics, flow-focusing junctions are used to generate highly monodisperse emulsions~\cite{Anna2003}. Thus, in Newtonian liquids, one typically observes two primary dynamical modes: dripping, in which monodisperse droplets are created inside the orifice, and jetting, in which the focused phase flows in parallel with the focusing phase beyond the orifice, only to break-up much later \cite{Baroud2010, Utada2007, Anna2003} due to the Rayleigh-Plateau instability \cite{Baroud2010}, although other regimes have also been reported \cite{Anna2003, Cubaud2008, Kovalchuk2019}.

\begin{figure}[t]
\includegraphics[width = 3.4 in, height = 1.5 in]{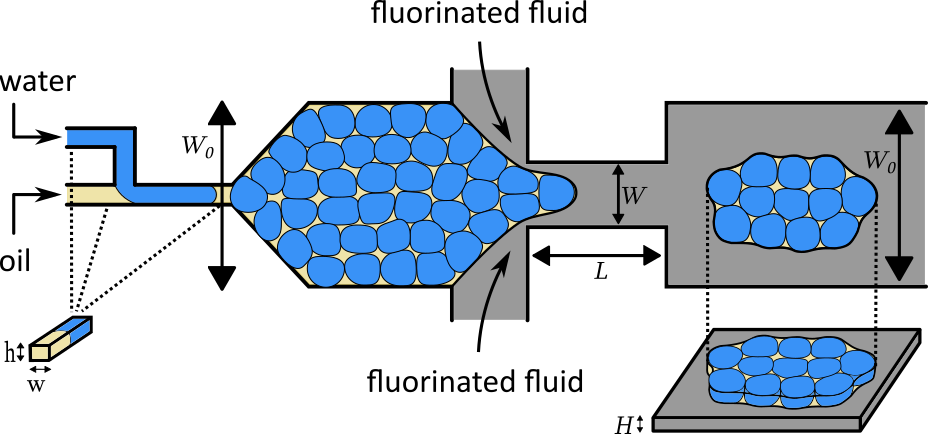}
	\caption{
	Scheme of the microfluidic system for generation of a quasi-2D granular medium (water-in-oil emulsion) and its fragmentation under flow-focusing with an external third immiscible phase (fluorinated fluid). Dimensions as measured by profilometry are: $W_0 = 2$ mm, $W = 1$ mm, $L = 2$ mm, $H = 0.11$ mm, $w = 0.11$ mm, $h = 0.08$ mm.
	\label{fig_scheme}}
\end{figure}

Here, in the case of a granular medium, we find new dynamical patterns, distinct from simple viscous jetting and dripping, such as (i) formation of fluctuating jets in which the fluctuations of jet width are influenced by avalanche-like 'discharge' of the dispersed granular phase at the junction rather than by the Rayleigh-Plateau instability of the jet, (ii) formation of very thin jets- single-file chains of grains--via 'unfolding' of thicker jets under extensional viscous stresses, and (iii) irregular break-up of the jets resulting in highly polydisperse grain clusters with a non-Gaussian size distribution. We highlight the stochasticity of the transport of the close-packed emulsion through the orifice in the various regimes and the impact of the behavior of individual grains on the dynamics of the entire emulsion (e.g., its break-up). Furthermore, we perform ad-hoc numerical simulations based on a recently developed Lattice Boltzmann method for multicomponent fluids with near-contact interactions  \cite{Montessori2019,Montessori2019a} which reproduce the experimental findings. The simulations employ a perfectly monodisperse emulsion which demonstrates that the observed stochasticity of the system is intimately associated with the granular structure and not, in particular, with polydispersity of the droplets.

The droplets ('grains') of the innermost phase are reproducibly formed at a T-junction of channels of rectangular cross-sections. Subsequently, the monodisperse emulsion is pushed into a wider channel (see Fig. \ref{fig_scheme}) and focused by the continuous phase into an orifice. We use three Newtonian liquids to formulate the double emulsion: fluorinated fluid \cite{Holtze2008} as the continuous phase, oil with surfactant as the middle (lubricating) phase, and dyed water as the innermost 'grain' phase (see SM for details). The T-junction generates droplets at a volume fraction of $86\%$ (the highest possible for which the emulsion is monodisperse and stable). Based on measured frequency of generation of the aqueous droplets, we estimate droplet volume to be around $0.11$ pL which yields the diameter of an undeformed spherical droplet $D_0 = 0.28$ mm. Since the value of $D_0$ is larger than the channel height $H=0.11$ mm, the droplets are flattened by the lower and upper walls. Based on the measured apparent areas of the generated clusters we estimate the diameter of the flattened droplets $D_{||} = 0.37$ mm with a coefficient of variation CV$_{D_{||}} = 9.2\%$. 

The ensuing emulsion is stable enough to produce flows for several minutes, with only occasional coalescence of the aqueous droplets. 

Our LB simulations, performed using a fully three-dimensional Color-Gradient approach augmented with near-contact interactions \cite{Montessori2019,Montessori2019a}, use slightly different, but similar, parameters for the system and droplets. See SM and \cite{Montessori2021} for details on the simulation methods and implementation.

\begin{figure}[t]
	\centering\includegraphics[width = 3.4 in, height = 2 in]{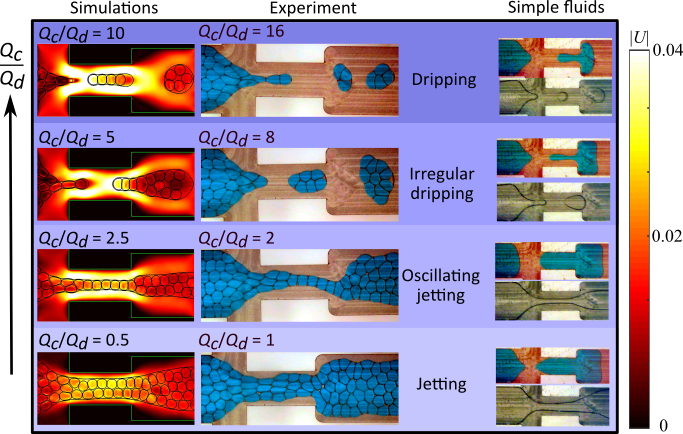}
	\caption{ 
Dynamical modes observed in the system upon varying $Q_c/Q_d$  $Q_d = 0.5$ mL/h. $Q_c/Q_d$ decreases from top to bottom (note different values for simulations and experiment). Smaller snapshots in the column on the right show flow patterns observed in the experiment with the emulsion replaced by a simple viscous liquid, either water (blue) or oil (transparent) for the same $Q_c/Q_d$ as in the experiment with the emulsion. The colorbar refers to simulation snapshots, with $U$ being the local velocity in lattice units. Experimental snapshots were taken from movies SM1 ($Q_c/Q_d = 1$), SM2 ($Q_c/Q_d = 2$), SM3 ($Q_c/Q_d = 8$), SM 4 ($Q_c/Q_d = 16$); simulation-based snapshots are taken from movies SM5 ($Q_c/Q_d = 0.5$), SM6 ($Q_c/Q_d = 2.5$), SM7 ($Q_c/Q_d = 5$), SM8 ($Q_c/Q_d = 10$)  
	\label{fig_phases}}
\end{figure}

In the experiment, we change the flow rate of the continuous phase $Q_c$, while keeping the flow rate of the dispersed phase (the emulsion) $Q_d$ constant and equal $Q_d = 0.5$ mL/h \footnote{at such $Q_d$ the emulsion was possibly most stable}. As a result, we observe several types of dynamic flow patterns as illustrated in Fig. \ref{fig_phases} and movies SM1-SM4. We find superficial similarity to jetting and dripping regimes present in simple fluids, however the observed dynamics is much richer. We can distinguish four different modes: (i) jetting with a large and moderately oscillating jet width, further referred to simply as \textit{jetting} (ii) jetting with thin, strongly oscillating jets and occasional break-up, further referred to as \textit{oscillating jetting} (iii) dripping resulting in a strongly polydisperse double-emulsion, further referred to as \textit{irregular dripping} (iv) dripping resulting in a relatively monodisperse double-emulsion, further referred to simply as \textit{dripping}. The numerical simulations recreate the same dynamical modes at values $Q_c/Q_d$ similar to yet slightly different than the experimental ones, see Fig. \ref{fig_phases} and movies SM5-SM8. 
We attribute the differences to slightly different geometrical parameters and volume fractions as imposed by the numerical constrains (see SM).

In order to understand the impact of granularity of the focused fluid on the onset of \textit{oscillating jetting}/\textit{irregular dripping} regimes, we repeat the flow focusing experiment with a simple fluid (water or oil) as the dispersed phase. We find simple jetting (see SM9 and SM10 for oil at $Q_c/Q_d$ = 1 and 2 respectively), highly monodisperse dripping (CV$_{A_{||}} = 2.2\%$ for oil, where $A_{||}$ is the area of a flattened oil drop, at $Q_c/Q_d$ = 3; see SM11) or bi-disperse dripping \cite{Garstecki2005} (the latter, with 2 narrow peaks, in the case with oil at high $Q_c/Q_d$, see SM12 and SM13 for examples with $Q_c/Q_d$ = 4 and 8 respectively); see SM for relevant histograms), but never observe irregular oscillations and rich dynamics similar to the case of a focused emulsion (see Fig. \ref{fig_phases}). In particular, we find only dripping for the case with water and the jetting-dripping transition for oil, which both agree with previous predictions for simple Newtonian liquids (i.e., that increasing viscosity of the dispersed phase promotes jetting) \cite{Cubaud2008}.

\begin{figure}[t]
	\centering\includegraphics[width = 3.4 in, height = 3.4 in]{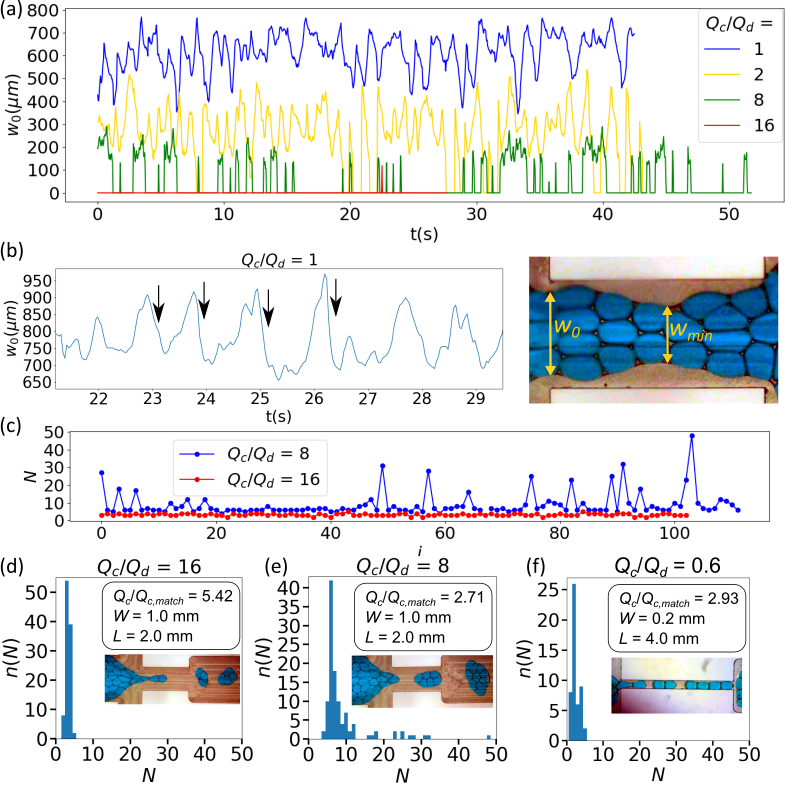}
	\caption{
	(a) Fluctuations of the minimum jet width $w_{min}(t)$
	(b) Fluctuations of jet width $w_0(t)$ at the entrance to the constriction. The avalanche-like events are marked with arrows. (c) Number of grains $N$ in clusters generated in the \textit{dripping} and \textit{irregular dripping} regimes ($Q_c/Q_d= 16$ and $8$) vs the index of cluster $i$ in order of generation, and (d)-(e) the corresponding histograms $n(N)$. The histogram in (f) shows analogous data for a system with a much thinner and longer orifice, yet with $Q_c/Q_{c,match}$ very close to the case in (e) (see main text and SM for further discussion).  
	\label{fig_statistics}}
\end{figure}

We further examine the four distinct dynamical modes observed for the focused emulsion in more detail. We define the minimum instantaneous jet width, $w_{min}(t) = \min_{x\in[0,L]} w(x,t)$ where $w(x,t)$ is the full spatio-temporal profile of the jet within the narrowing, as a measure of jet oscillations in time (Fig. \ref{fig_statistics}a). We find that the corresponding time average, $\langle w_{min} \rangle = T^{-1}\int_0^{T}\text{d}t\,w_{min}(t)$, where $T$ is the time of duration of the experiment, decreases upon increasing $Q_d/Q_c$ while the stochastic fluctuations of $w_{min}(t)$ remain of similar absolute magnitude. Accordingly, this leads to occasional break-up ($w_{min}=0$) of the jet in the oscillating jetting mode.

Additionally, in the \textit{jetting} mode, we frequently observe abrupt granular 'discharge' of the junction, associated with rapid entrance of several droplets in-parallel into the constriction. In order to quantify such avalanche-like behavior we measure the width of the jet $w_0(t)$ precisely at the entrance to the constriction. We find that $w_0(t)$ develops a saw-tooth like profile (Fig.\ \ref{fig_statistics}b) characteristic of avalanches and previously also observed in sheared foams and dense suspensions \cite{Durian1995, Kumar2020}.

Next, we measure the sizes of the subsequently generated clusters in the \textit{dripping} and \textit{irregular dripping} modes (Fig. \ref{fig_statistics}c). Whereas in the former case the clusters are relatively monodisperse (yet much more polydisperse than in dripping of simple viscous fluids), in the latter case we observe recurring peaks in the cluster size corresponding to extremely large clusters. More quantitatively, in the \textit{dripping} mode the number of grains $N$ in a cluster does not apparently deviate from the Gaussian distribution. The coefficient of variation CV$_N = 19.5\%$ (see Fig. \ref{fig_statistics}d) is significantly larger then in the case with the granular emulsion replaced by the pure oil phase (CV$_{A_{||}} = 2.2\%$), yet still moderate. In contrast, in the \textit{irregular dripping} mode the distribution of cluster sizes $N$ features a long right tail for large $N$ (see Fig. \ref{fig_statistics}e), with CV$_N =74.3\%$ and a very large skewness, as documented by a Pearson's moment coefficient of skewness (see SM for a formal definition) $S_N$ = 3.1.

We associate the formation of the extremely large clusters in the irregular dripping regime with the emergence of single-file chains of grains within the narrowing, which, once formed, exhibit remarkable stability. In principle, such chains remain stable once the 
local velocity of the continuous phase around the chain $U_c$ matches the 
velocity of the grains inside the chain $U_{d,chain}$, which in turn is set by 
the rate of feeding of the grains into the orifice (note that this condition also determines the boundary between the dripping and jetting modes). 
Considering that $U_c = Q_c/[H \times (W-W_{chain})]$ and $U_{d,chain} = Q_d/(H\times W_{chain})$, where $W_{chain}$ is the width of the chain, the requirement 
$U_c = U_{d,chain}$ leads to the following condition on the matching flow rate of the continuous fluid $Q_{c,match}$:

\begin{equation}\label{eq:the_one_and_only}
    Q_{c,match}/Q_d =W/W_{chain}-1
\end{equation}

\begin{figure}[t]
	\centering\includegraphics[width = 3.4 in, height = 3.8 in]{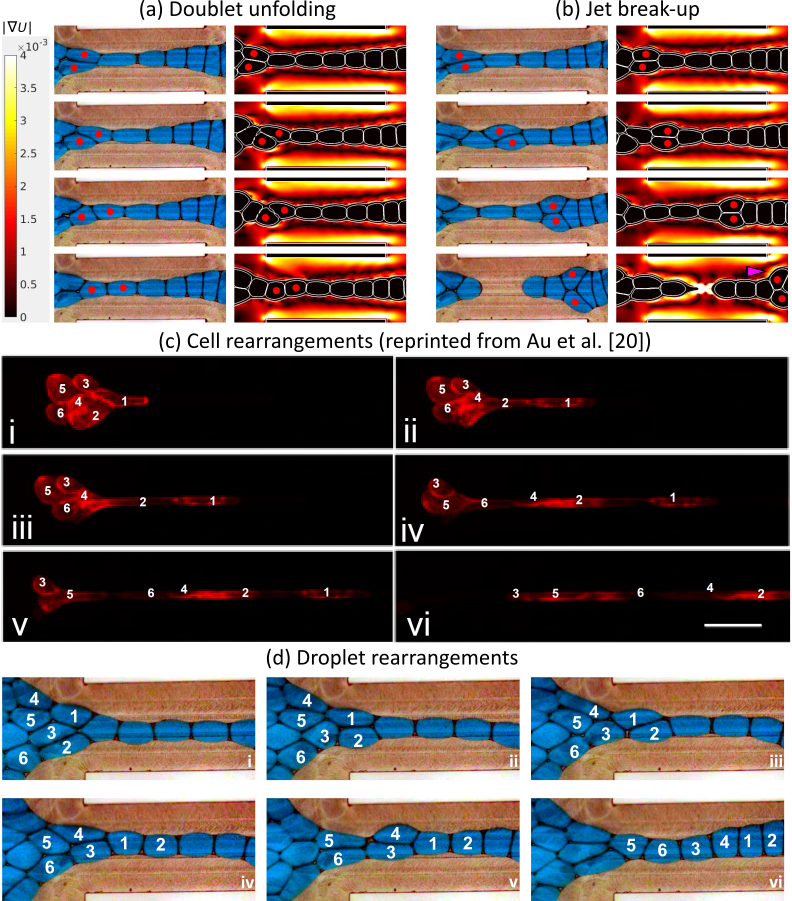}
	\caption{
	(a)-(b) Numerical (at $Q_c/Q_d$ = 3.5) and experimental (at $Q_c/Q_d$ = 8) snapshots (see SM14 and SM15 respectively for the full movies) visualizing (a) doublet unfolding upon entry into the constriction and (b) a failure of unfolding, leading to an increased velocity gradient $\nabla U$ around the doublet (pink marker in the simulation panel) and  break-up of the jet. (c) The order of entry of cells forming a CTC cluster in a narrowing channel under co-flow (figure adapted with permission from \citet{Au2016}; copyright National Academy of Sciences 2016) and (d) the order of entry of droplets into the constriction in our experiment, at $Q_c/Q_d = 8$ (see SM16 for the full movie). 
	\label{fig_seq_n4}}
\end{figure}

\noindent We note that this requirement resembles the condition for continuity of soft polymer fibers stretched by an accelerating co-flow, studied by \citet{Mercader2010}. In fact, our granular chains resemble semi-solid fibers rather than viscous jets as demonstrated by (i) the lack of the Rayleigh-Plateau instability (typical of viscous jets) \cite{Utada2007} and (ii) longitudinal stretching and/or compression of the chain as visualized by droplet deformations within the constriction, see Fig. \ref{fig_seq_n4}a. We associate such elastic solid-like behavior with a combination of the capillary arrest and deformability of the droplets within the chain. 

From the experimentally measured average width of chain $W_{chain} = 0.68 D_{||} = 0.253\, W$ we obtain $Q_{c,match}/Q_d = 2.95$. This is close to the value $Q_c/Q_d = 2$ corresponding to the \textit{oscillating jetting} regime; however, we actually observe single-file chains more often when $Q_c/Q_d = 8$, in the \textit{irregular dripping} mode. We suspect that the relative scarcity of single-file chains for $Q_c/Q_d = 2$ results from spontaneous 'folding' of single-file chains into wider jets in the immediate vicinity of the theoretical matching velocity $U_{d,chain}$ (see SM2, frames 51-72, 145-164, 670-691). At the same time, due to the finite length of the orifice $L$, chains are able to survive extensional stresses at $Q_c \gtrsim Q_{c,match}$ which may explain their abundance at  $Q_c/Q_d = 8$.

We note that even at matched velocities the chains can break due to irregularity of the grain feeding into the orifice associated with stochasticity of grain rearrangements upon approaching the constriction. When a pair of grains enter the narrowing simultaneously they may rearrange, or 'unfold', into a chain or not- in the latter case entering as a 2-grain cluster or a 'fold' (see Fig. \ref{fig_seq_n4}). When the fold enters the orifice the continuous phase needs to locally accelerate and pass around it to conserve flux. The increased viscous forces acting at the fold result in chain stretching via longitudinal grain deformation which may eventually cause chain breakup. 

Our LB simulations allow us to extract precise information about the velocity gradients within the system and verify this scenario. Indeed, we find progressively increasing velocity gradients around the doublet (see the purple marker in the last snapshot of Fig. \ref{fig_seq_n4}b). We propose that a similar mechanism (i.e., the acceleration of the continuous phase around wider parts of the jet) might also lead to the enhancement of fluctuations of jet width in the \textit{jetting} and the \textit{oscillating jetting} modes.

Next, we also perform a series of experiments with a smaller width of the orifice ($W \lesssim D_{||}$), for which a simultaneous 
entry of two grains into the narrowing is hindered (see SM for details). In this case, the complex dynamical picture is lost and we only observe a transition between single-file jetting and dripping. To provide an example, we quantify the cluster size distribution in this geometry in Fig. \ref{fig_statistics}f  when $Q_c/Q_{c,match}$ (which serves as a measure of proximity to the jetting-dripping transition) is almost identical as in the long-tailed \textit{irregular dripping} mode (Fig. \ref{fig_statistics}e). We still observe strong polydispersity (CV$_{N}= 44 \%$), however, no long tails ($S_{N} = 0.73$).  
This further confirms the impact of individual grain rearrangements and, more specifically, the manner in which the 
grains enter the constriction, on the fate of the entire system, including the large-scale stochastic behavior.

Finally, we provide an example of how granular rearrangements observed in our flow-focusing setup could serve as a 'benchmark' for more complex soft granular flows including confined biological flows. In fact, sequences of cell rearrangements have been previously studied in circulating tumor cell clusters transiting a narrowing channel \cite{Au2016}. Upon approaching a constriction the clusters were often able to unfold into a single-file chain without break-up. In some cases, the order in which the cells approached the narrowing seemed to determine their order of entry, but in some other cases the order was strongly disturbed by cell rearrangements (see Fig. \ref{fig_seq_n4}c). This is interpreted in \cite{Au2016} as the effect of heterogeneity of cell-cell interactions and polydispersity of cells. 
However, our experiments demonstrate that even in a homogeneous, monodisperse \textit{passive} granular system the order of entry is not strictly determined by the order of approach but rather depends on stochastic rearrangements upon entry (see Fig. \ref{fig_seq_n4}d). 
Accordingly, we argue that the phenomena reported by \citet{Au2016} may result from the 
immanent irregularity of flow patterns associated with many-body interactions and general 
stochastic dynamics of soft granular media, and not only from heterogeneity of the grains.

In summary, we develop a model platform to study the behaviour of soft granular media subjected to external flows and demonstrate rich phenomenology including stochastic granular jetting- and dripping-like modes with no counterpart in simple fluids.

We note that series of two (or more) microfluidic junctions have been previously used to produce double-emulsion core-shell droplets with multiple cores \cite{Okushima2004, Lee2009, Wan2008, Abate2009, Adams2012, Utada2005, Vladisavljevic2017}. A couple of recent works considered cores-in-shell volume fractions high enough ($>80\%$) for the double-emulsion drops to be considered soft granular clusters. \cite{Guzowski2015, Constantini2018, Kim2011, Lee2009}. However, those previous works exploited generation of the clusters via one-by-one  feeding of the cores into the shell without actually considering the flow of a soft-granular medium \textit{per se}. In this Letter, we argue that the latter poses a completely different problem and involves phenomena not present in simple fluids.

Our findings open up several avenues for future work. First, the full dynamical phase diagram in the 3-dimensional $(Q_c,Q_d,\phi)$-space including possible hysteretic behavior at transitions between the modes--also depending on the viscosities and interfacial tensions--remains to be established. Second, the statistics of rearrangements between individual grains could be further investigated to shed light on the effective phases of matter (solid- vs fluid-like) occurring in such a system. Finally, our platform could also be further developed to allow tracking of the internal relaxation dynamics of the generated granular clusters. This poses possible significance e.g., to the recovery of tissues after mechanical injury or the dynamics of CTC's in capillaries during cancer metastasis. 
\\

The authors acknowledge funding from the European Research
Council under the European Union’s Horizon 2020 Framework Programme (Grant No. FP/2014–2020), ERC Grant Agreement No.
739964 (COPMAT), Marie Sk\l odowska-Curie grant No. 847413 and the PRACE 16DECI0017 RADOBI project. M.B. acknowledges the PMW 
programme of the Minister of Science and Higher Education in the years 2020-2024 no. 5005/H2020-MSCA-COFUND/2019/2.
A.M. acknowledges the CINECA Computational Grant ISCRA-C IsC83 - “SDROMOL”, id. HP10CZXK6R 
under the ISCRA initiative, for the availability of high performance computing resources needed to run the simulations and the support provided.
J.G. acknowledges support from Foundation for Polish Science within First Team program under grant no POIR.04.04.00-00-26C7/16-00. The authors thank PRL reviewers for insightful comments which helped to significantly improve the quality of the manuscript. M.B. and J.G. thank Patryk Adamczuk and Mikołaj Boroński for technical assistance.

\bibliography{main}

\end{document}


\preprint{APS/123-QED}

\title{Supplemental Material for the article \textit{Stochastic jetting and dripping in confined soft granular flows}}

\author{Micha{\l} Bogdan}
\email[]{mbogdan@ichf.edu.pl}
\affiliation{Institute of Physical Chemistry, Polish Academy of Sciences, Kasprzaka 44/52,
01-224 Warsaw, Poland}

\author{Andrea Montessori}
\affiliation{Dipartimento di Ingegneria, Università degli Studi Roma tre, via Vito Volterra 62, Rome, 00146, Italy}

\author{Adriano Tiribocchi}
\affiliation{Istituto per le Applicazioni del Calcolo del Consiglio Nazionale delle Ricerche, via dei Taurini 19, 00185, Rome, Italy}

\author{Fabio Bonaccorso}
\affiliation{Istituto per le Applicazioni del Calcolo del Consiglio Nazionale delle Ricerche, via dei Taurini 19, 00185, Rome, Italy}
\affiliation{Department of Physics and National Institute for Nuclear Physics, University of Rome "Tor Vergata'', Via Cracovia, 50, 00133 Rome, Italy}

\author{Marco Lauricella}
\affiliation{Istituto per le Applicazioni del Calcolo del Consiglio Nazionale delle Ricerche, via dei Taurini 19, 00185, Rome, Italy}

\author{Leon Jurkiewicz}
\affiliation{Institute of Physical Chemistry, Polish Academy of Sciences, Kasprzaka 44/52,
01-224 Warsaw, Poland}

\author{Sauro Succi}
\affiliation{Istituto per le Applicazioni del Calcolo del Consiglio Nazionale delle Ricerche, via dei Taurini 19, 00185, Rome, Italy}
\affiliation{Center for Life Nanoscience at la Sapienza, Istituto Italiano di Tecnologia, viale Regina Elena 295, 00161, Rome, Italy}
\affiliation{Department of Physics, Harvard University, 17 Oxford St, Cambridge, MA 02138, United States}

\author{Jan Guzowski}
\email[]{jguzowski@ichf.edu.pl}
\affiliation{Institute of Physical Chemistry, Polish Academy of Sciences, Kasprzaka 44/52,
01-224 Warsaw, Poland}

\date{\today}

\maketitle


\section{\label{sec:materials_and_methods} Materials and methods}

All microfluidic channels are fabricated via milling in polycarbonate. Precise parameters of the Newtonian liquids used to formulate the double emulsion: fluorinated fluid FC40 with $1\%$ w/w fluorosurfactant (PFPE-PEG-PFPE \cite{Holtze2008}) as the continuous phase, a mixture of silicone oil (PMX 200, 5cSt), hexadecane, and the surfactant SPAN80 in w/w proportions 70:30:1 as the middle (lubricating) phase (referred in the main text and further here as 'oil'), and water dyed with 0.1\% w/w Erioglaucine as the innermost 'grain' phase. Based on literature values, we estimate the corresponing viscosities to be $4.1$, $4.0$, and $3.4$ mPa s, respectively. The interfacial tensions of the water-oil and oil-external interfaces are $3.4$ and $4.9$ mN/m, respectively, as measured by pendant drop method.

In LB simulations, the geometrical parameters, expressed in simulation units, are as follows: $W_0=L=280$, $W=140$, $H=30$, while the total length of the computational domain is $846$. The droplets are generated by imposing an internal boundary condition, as proposed in \cite{Montessori2021}; the volume fraction is estimated as $\sim0.75$, with a completely uniform droplet diameter of $D =45$ simulation units.

\section{\label{sec:alternative set-up} Formal definition of Pearson's coefficient of skewness}

 The formal definition of the Pearson's moment coefficient of skewness $S_X$ for a random variable $X$ (which we use to quantitatively assess the asymmetry of the measured probability distributions of sizes of granular clusters) is: .
 
 \begin{equation}
     S_X = E\left[\left( \frac{X-\mu_X}{\sigma_X} \right)^3\right]
 \end{equation}
 
 Where $E$ is the expectation value of the expression in the squared parenthesis, while $\mu_X$ and $\sigma_X$ are respectively the mean and standard deviation of the variable $X$.  

\section{\label{sec:alternative set-up} Experiments with alternative system dimensions}

\begin{figure}[ht]
\includegraphics[width = 3.4 in]{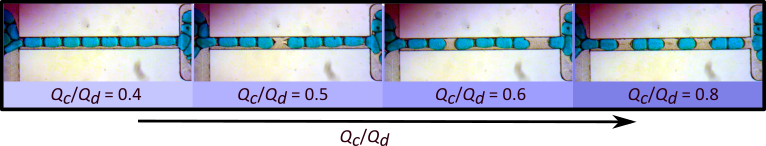}
	\caption{
	 Behavior of the emulsion upon passing a very narrow constriction under varying $Q_c/Q_d$.
	\label{fig_alternative_set_up}}
\end{figure}

In order to further verify the role of the formation of 2-grain 'folds' upon entry into the orifice in the behavior of the system, we repeat the experiments in an alternative set-up, in which the orifice is narrow enough to prevent the formation of such folds. The microfluidic system is again as in Fig. 1 of the main text, but with the following dimensions modified to the values specified: $W = 0.2$ mm, $L = 4$ mm, $H = 0.18$ mm. Note this means the orifice is now narrower than the diameter of a single grain. As expected, the simultaneous entry of two droplets into the narrowing is now impossible (see Fig. \ref{fig_alternative_set_up} and supporting movies, respectively SM17 for $Q_c/Q_d$=0.4, SM18 for $Q_c/Q_d$=0.5, SM19 for $Q_c/Q_d$=0.6, SM20 for $Q_c/Q_d$=0.8). The complex dynamical picture as described in the main text is now lost, with long right tails characteristic of the \textit{irregular dripping} regime absent; we now observe a simple transition between jetting and dripping upon increasing $Q_c/Q_d$, with a reduction of cluster size within the dripping regime upon a further increase of $Q_c/Q_d$. 

\section{\label{sec:pure_oil} Flow focusing of pure 'oil'}

\begin{figure}[ht]
\includegraphics[width = 3.4 in]{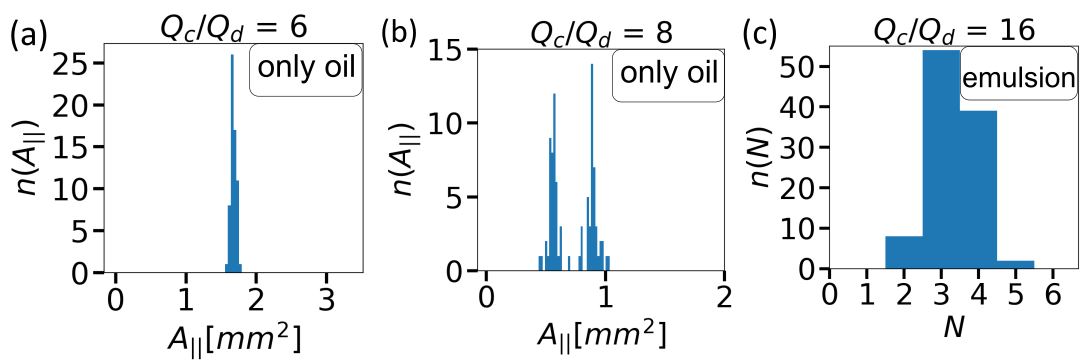}
	\caption{
	 Histograms of numbers of grains $N$ in clusters of the emulsion $n(N)$ and of areas $A_{||}$ of flattened drops of the oil $n(A_{||})$, formed during dripping of the emulsion and the pure oil respectively.
	\label{fig_oil}}
\end{figure}

In Fig. \ref{fig_oil} we illustrate the effect of granularity on the variation of the areas of formed clusters/drops by comparing histograms of $N$ for the emulsion's granular \textit{dripping} mode (panel c) with histograms of $A_{||}$ for monodisperse (panel a) and bi-disperse (panel b) dripping modes of the oil itself. The $x$ axes are re-normalised to the mean of the distribution in each case to help visualise relative variations.

\bibliography{SM}